\documentclass[pra,twocolumn,superscriptaddress,showpacs,preprintnumbers,amsmath,amssymb]{revtex4}
\usepackage{mathrsfs}
\usepackage{bbm}
\usepackage{amsfonts}
\usepackage{tipa}

\usepackage{epsfig,graphicx}
\usepackage{amstext}
\usepackage{amsmath}
\usepackage{graphicx}
\usepackage{times}

\begin{document}


\title{Preservation of the geometric quantum discord in noisy environments}

\author{Ming-Liang Hu}
\email{mingliang0301@163.com}
\affiliation{School of Science, Xi'an Jiaotong University,
             Xi'an 710049, China}
\affiliation{School of Science, Xi'an University of Posts and
             Telecommunications, Xi'an 710121, China}
\author{Dong-Ping Tian}
\affiliation{School of Science, Xi'an Jiaotong University,
             Xi'an 710049, China}

\begin{abstract}
Geometric description of quantum correlations are favored for their
distinct physical significance. Geometric discord based on the trace
distance and the Bures distance are shown to be well-defined quantum
correlation measures. Here, we examine their particular dynamical
behaviors under independent as well as common structured reservoirs,
and reveal their robustness against decoherence. We showed that the
two well-defined geometric discord may be preserved well, or even be
improved and generated by the noisy process of the common reservoir.
Moreover, we also provided a strategy for long-time preservation of
these two geometric discord in independent reservoirs.
\end{abstract}


\maketitle

\section{Introduction}\label{sec:1}
The existence of quantum correlations in a system is one of the most
remarkable features of quantum theory which differentiates the
quantum world from that of the classical one, and quantifying and
understanding quantum correlations remains the subject of active
research since the early days of quantum mechanics \cite{Horodecki}.
In the past two decades, a broad survey of different aspects of
quantum correlations, such as the Bell-type correlations
\cite{Bell}, the capacity for teleportation \cite{Bennett}, and a
plethora of measures for quantum entanglement \cite{Horodecki}, were
performed. Particularly, since the pioneering work of Ollivier and
Zurek \cite{Ollivier}, and that of Henderson and Vedral
\cite{Henderson}, the concept of quantum discord (QD) as a more
general quantum correlation measure than that of entanglement,
prompted a huge surge of people's research interest from different
perspectives, see Refs. \cite{Modirmp,IJTP} for a comprehensive
review.

Originally, the QD was defined through the discrepancy between two
expressions of the mutual information that are classically identical
and quantum-mechanically inequivalent \cite{Ollivier}. This is
indeed an entropic measure of quantum correlation, and was favored
for its operational interpretations \cite{int1,int2,int3,Gumile} and
potential applications in various quantum tasks
\cite{Datta,Lanyon,Werlang}. But its evaluation is very hard due to
the optimization procedure involved, and the closed expressions are
known only for certain special (such as the Bell-diagonal
\cite{luos1}) states. Particularly, it has been shown that
analytical evaluation of QD for general states is impossible
\cite{Girolami}. Therefore, other measures of quantum correlations
which are easy to calculate are needed. In this respect, Luo
presented the concept of measurement-induced disturbance
\cite{luos2}, where the measurement is induced by the spectral
resolutions of the reduced states of a system. As such, it evades
the procedure of optimization which is usually intractable.

Another routine for characterizing quantum correlations is via the
geometric approach based on different distance measures. The seminal
work along this line was that accomplished by Daki\'{c} {\it et al.}
\cite{Dakic}. They proposed to use the square of the minimal
Hilbert-Schmidt distance as a basis for defining QD, and
subsequently, Luo and Fu presented a variational and equivalent
definition for it based on von Neumann measurements, and derived a
tight lower bound for general bipartite states \cite{Luo}. The
figure of merit for this geometric measure of QD lies in its
analytical evaluation for general two-qubit states \cite{Dakic} and
certain bipartite states with high symmetry \cite{Luo,Chitambar}. It
also plays a crucial role in specific quantum protocols, such as
remote state preparation \cite{Dakicnp}. But this measure of
geometric discord may be increased by trivial local operations on
the unmeasured subsystem \cite{Piani}, and thus was not a good
measure of quantum correlations \cite{Henderson}. To avoid this
shortcoming, geometric discord based on other distance measures were
proposed, e.g., the modified version of the geometric discord
defined by making use of the square root of the density operator
\cite{luos3}. Here, we will consider the geometric discord defined
by employing the trace distance \cite{trace} and the Bures distance
\cite{bures}. This way of characterizing quantum correlation has
previously been suggested by Luo and Fu in their pursuing analytical
solutions for the geometric discord \cite{Luo}, and has been
exploited explicitly very recently \cite{trace,bures,Montealegre}.
They can circumvent the problem occurs for the geometric discord
defined in Ref. \cite{Dakic}, and therefore can be regarded as
well-defined measures of quantum correlations.

From a practical point of view, one may wonder its robustness
against decoherence after the introduction of a new well-defined
quantum correlation measure. Due to the advantage of the distance
measures adopted for defining quantum correlations, it is expected
that the aforementioned two geometric discord will exhibit different
behaviors under decoherence \cite{Montealegre}, and a comparative
study of this issue may provide us with information that is
essential to various quantum protocols, particularly those based
only on them.

In this paper, we take an investigation of the above problem. To be
explicitly, we consider robustness of the foregoing two well-defined
geometric discord for a central two-qubit system coupled to noisy
environments. We will compare their particular dynamical behaviors,
and try to provide effective methods for fighting against the
deterioration of them, as this is of special importance to various
quantum protocols.

\section{Well-defined measures of geometric discord}\label{sec:2}
To begin with, we first briefly review the definitions as well as
the general formalism for the trace distance and the Bures distance
geometric discord. For a bipartite system $AB$ described by the
density operator $\rho$, the trace distance discord is defined as
the minimal trace distance between $\rho$ and all of the
classical-quantum states $\rho_{CQ}$ \cite{trace}, namely,
\begin{eqnarray}\label{eq1}
 D_{\rm T}(\rho)=\min_{\chi\in\rho_{CQ}}||\rho-\chi||_1,
\end{eqnarray}
where $||X||_1={\rm Tr}\sqrt{X^\dag X}$ denotes the trace norm
(Schatten 1-norm), and $\rho_{CQ}$ takes the following form
\begin{eqnarray}\label{eq2}
 \rho_{CQ}=\sum_i p_i \Pi_i^A\otimes \rho_i^B,
\end{eqnarray}
which is a linear combination of the tensor products of $\Pi_i^A$
(the orthogonal projector in the Hilbert space $\mathcal {H}_A$) and
$\rho_i^B$ (an arbitrary density operator in $\mathcal {H}_B$), with
$\{p_i\}$ being a probability distribution.

For the special case of the two-qubit \emph{X} states $\rho^X$ whose
possible nonzero elements are along only the main diagonal and
anti-diagonal \cite{xstate}, the trace distance discord can be
derived analytically \cite{analytic}, which is of the following
compact form
\begin{eqnarray}\label{eq3}
 D_{\rm T}(\rho^X)=\sqrt{\frac{\gamma_1^2
 \gamma_{\rm max}^2-\gamma_2^2\gamma_{\rm min}^2}
 {\gamma_{\rm max}^2-\gamma_{\rm min}^2+\gamma_1^2-\gamma_2^2}},
\end{eqnarray}
where $\gamma_{1,2}=2(|\rho_{23}|\pm |\rho_{14}|)$,
$\gamma_3=1-2(\rho_{22}+\rho_{33})$, $\gamma_{\rm
max}^2=\max\{\gamma_3^2,\gamma_2^2+x_{A3}^2\}$, and $\gamma_{\rm
min}^2=\min\{\gamma_1^2,\gamma_3^2\}$, with
$x_{A3}=2(\rho_{11}+\rho_{22})-1$.

If one further consider a specific subset of the \emph{X} states,
i.e., the Bell diagonal states of the form $\rho^{\rm
BD}=\frac{1}{4}[I_2\otimes I_2+\vec{c}\cdot
(\vec{\sigma}\otimes\vec{\sigma})]$, with $\vec{c}=\{c_1,c_2,c_3\}$
being a three-dimensional vector with elements satisfying $0
\leqslant |c_i| \leqslant 1$, and
$\vec{\sigma}=\{\sigma_1,\sigma_2,\sigma_3\}$ denotes the standard
Pauli matrices, the trace distance discord can be further simplified
as \cite{trace}
\begin{eqnarray}\label{eq4}
 D_{\rm T}(\rho^{\rm BD})={\rm int}\{|c_1|,|c_2|,|c_3|\},
\end{eqnarray}
which is in fact the intermediate value for the absolute values of
the correlation functions $c_1$, $c_2$, and $c_3$.

Different from that of the trace distance discord, the Bures
distance geometric discord is defined via the Bures distance
$d_{\text{B}}(\rho,\sigma)=2[1-\sqrt{F(\rho,\sigma)}]$ between two
density operators $\rho$ and $\sigma$ \cite{bures}, which is similar
with that of the Bures measure of entanglement \cite{bentangle}.
Here, we take the definition of Ref. \cite{bures2}, which is of the
following form
\begin{eqnarray}\label{eq5}
 D_{\rm B}(\rho)=\sqrt{(2+\sqrt{2})[1-\sqrt{F_{\rm max}(\rho)}]},
\end{eqnarray}
where $F_{\rm max}(\rho)=\max_{\chi\in\rho_{CQ}}F(\rho,\chi)$
denotes the maximum of the Uhlmann fidelity $F(\rho,\chi)=[{\rm
Tr}(\sqrt{\rho}\chi\sqrt{\rho})^{1/2}]^2$. Note that $D_{\rm
B}(\rho)$ in Eq. \eqref{eq5} is normalized, and its square equals to
that defined in Ref. \cite{bures}.

There are several special cases that the evaluation of the Bures
distance discord can be simplified. (i) Pure state $|\Psi\rangle$.
For this case we have $F_{\rm max}(|\Psi\rangle)=\mu_{\rm max}$,
with $\mu_{\rm max}$ being the largest Schmidt coefficient of
$|\Psi\rangle$ \cite{bures}. (ii) The Bell-diagonal states
$\rho^{\rm BD}$, for which we have \cite{bures2,bures3}
\begin{eqnarray}\label{eq6}
 F_{\rm max}(\rho^{\rm BD})&=&\frac{1}{2}+\frac{1}{4}\max_{\langle ijk\rangle}
 \bigg[\sqrt{(1+c_i)^2-(c_j-c_k)^2} \nonumber\\
 &&+\sqrt{(1-c_i)^2-(c_j+c_k)^2}\bigg],
\end{eqnarray}
where the maximum is taken over all the cyclic permutations of
$\{1,2,3\}$. (iii) For the $2\times n$-dimensional system, although
there is no analytic solution, the maximum of the Uhlmann fidelity
can be calculated as \cite{bures3}
\begin{eqnarray}\label{eq7}
 F_{\rm max}(\rho)=\frac{1}{2}\max_{||\vec{u}=1||}
                   \left(1-{\rm Tr}\Lambda(\vec{u})
                   +2\sum_{k=1}^{n_B}\lambda_k(\vec{u})\right),
\end{eqnarray}
where $\lambda_k(\vec{u})$ represents the eigenvalues of
$\Lambda(\vec{u})=\sqrt{\rho}(\sigma_{\vec{u}}\otimes I_{n_B})
\sqrt{\rho}$ in non-increasing order, and
$\sigma_{\vec{u}}=\vec{u}\cdot\vec{\sigma}$ with
$\vec{u}=(\sin\theta\cos\phi,\sin\theta\sin\phi,\cos\theta)$ being a
unit vector in $\mathbb{R}^3$, and $n_B$ the dimension of $\mathcal
{H}_B$.

\section{The model}\label{sec:3}
After recalling the basic formalism for the trace distance and the
Bures distance geometric discord, we now present the model for our
system and the scenario of system-environment coupling. The central
system we considered consists of two identical qubits, and they are
subject to either of the following two representative structured
reservoirs: (i) the independent or (ii) the common zero-temperature
reservoir \cite{Breuer}. The corresponding Hamiltonian are given
respectively by
\begin{eqnarray}\label{eq8}
 \hat{H}_{i}=\omega_0\sum_{n}\sigma^n_{+}\sigma^n_{-}
           +\sum_{k,n}(\omega_k^n b_k^{n\dag} b_k^n
           +g_k^n b_k^n\sigma^n_{+}+{\rm h.c.}),
\end{eqnarray}
and
\begin{eqnarray}\label{eq9}
 \hat{H}_{c}=\omega_0\sum_n\sigma^n_{+}\sigma^n_{-}
           +\sum_{k}\omega_k b_k^{\dag} b_k+\sum_{k,n}(g_k b_k
           \sigma^n_{+}+{\rm h.c.}),
\end{eqnarray}
where $\omega_0$ and $\omega_k$ denote, respectively, the transition
frequency of the two qubits and frequency of the reservoir field
mode $k$ with the bosonic creation (annihilation) operator
$b_k^{\dagger}$ ($b_k$) and the system-reservoir coupling constant
$g_k$. Moreover, $\sigma_\pm=(\sigma_1 \pm i\sigma_2)/2$, and the
superscript $n$ in the summation runs over the two qubits and their
respective reservoir. Note that we considered here that the two
qubits are sufficiently separated from each other and, therefore, no
direct interactions between them have been taken into account.

The above model was used to study dynamics of entanglement
\cite{indep,common,common1,Bellomo}, entropic discord
\cite{discord1,discord2,discord3}, and other related topics
\cite{Yu,Sun,Hu}. The evolution of the two qubits under the
system-environment coupling depends on the particular choice of the
spectral density of the reservoir. In this paper, we take the
Lorentzian spectral distribution of the following form \cite{Breuer}
\begin{eqnarray}\label{eq10}
 J(\omega)=\frac{1}{2\pi}\frac{\gamma_0\lambda^2}{(\omega-\omega_0)^2+\lambda^2},
\end{eqnarray}
where the parameters $\lambda$ and $\gamma_0$ define the spectral
width of the reservoir and the decay rate, respectively. They are
associated with the reservoir correlation time $\tau_B$ and the
relaxation time $\tau_R$ by $\tau_B\approx\lambda^{-1}$ and
$\tau_R\approx\gamma_0^{-1}$, and their relative magnitudes
determine the Markovian ($\lambda>2\gamma_0$) and the non-Markovian
($\lambda<2\gamma_0$) regimes.

For the above scenario of system-environment coupling, the
evaluation of the time-evolved density matrix $\rho(t)$ has already
been discussed in the literature \cite{Breuer}. Here, we point out
that for the independent reservoir, analytical solutions of
$\rho(t)$ can be derived for arbitrary initial states \cite{indep},
while for the common reservoir, $\rho(t)$ can be solved numerically
via the pseudomode approach \cite{common1}, and for the special case
of the initial two-qubit extended Werner-like states, compact form
of $\rho(t)$ can also be obtained by using the technique of Laplace
transformation \cite{common2}.

\section{Robustness and preservation of the geometric discord}\label{sec:4}
With the help of the above preliminaries, we now begin our
discussion about robustness of the geometric discord $D_{\rm
T}(\rho)$ and $D_{\rm B}(\rho)$ under the Lorentzian structured
reservoir. We will take $|\Phi\rangle=\alpha|10\rangle +
\sqrt{1-\alpha^2}|01\rangle$ as the initial state of the two qubits,
and for the sake of simplicity, we will do not list the explicit
form of $\rho(t)$ here as they can be easily written via the methods
mentioned above  \cite{indep,common2}.
\begin{figure}
\centering
\resizebox{0.40\textwidth}{!}{%
\includegraphics{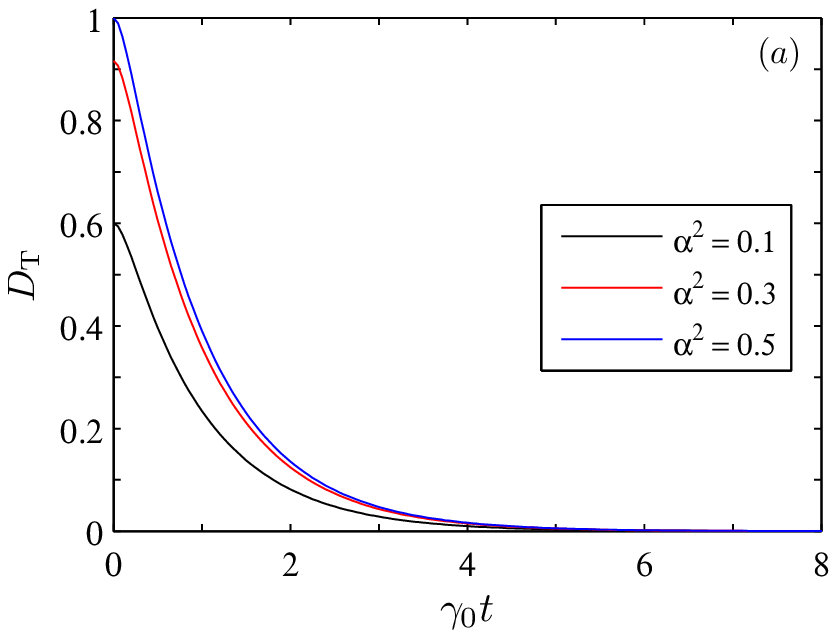}}
\centering
\resizebox{0.40\textwidth}{!}{%
\includegraphics{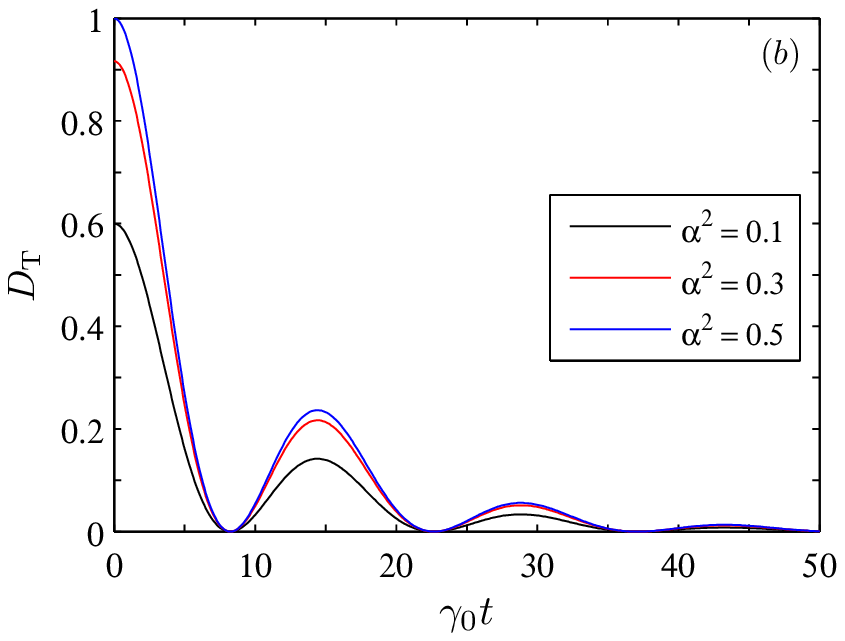}}
\caption{(Color online) Trace distance discord $D_{\rm T}(\rho)$
         versus $\gamma_0 t$ for the initial states $|\Phi\rangle$
         in independent reservoirs, where the parameter
         $\lambda$ is chosen to be $\lambda=10\gamma_0$ (a) and
         $\lambda=0.1\gamma_0$ (b), respectively.}
         \label{fig:1}
\end{figure}

When considering the trace distance discord $D_{\rm T}(\rho)$ for
the initial state $|\Phi\rangle$, our calculation shows that it is a
symmetric quantity with respect to $\alpha^2=0.5$. In Fig.
\ref{fig:1}, we plotted dynamics of $D_{\rm T}(\rho)$ versus the
scaled time $\gamma_0 t$ for the case of the two qubits subject to
independent reservoirs, from which one can note that $D_{\rm
T}(\rho)$ takes its maximum at $\alpha^2=0.5$, and decreases with
the increase of $|\alpha^2-0.5|$. For fixed $\alpha^2$, $D_{\rm
T}(\rho)$ decays monotonically with increasing $\gamma_0 t$ in the
Markovian regime [Fig. \ref{fig:1}(a)], while it exhibits damped
oscillations in the non-Markovian regime [Fig. \ref{fig:1}(b)], and
suffers instantaneous disappearance at the critical time
$t_n=2[n\pi-\arctan(d/\lambda)]/d$, with
$d=\sqrt{2\gamma_0\lambda-\lambda^2}$ and $n\in\mathbb{Z}$
\cite{indep}. As there are no direct interactions between the two
qubits, and the two qubits interact respectively with their own
independent reservoir, the revivals of $D_{\rm T}(\rho)$ after its
instantaneous disappearance in the non-Markovian regime is induced
by the memory effects of the reservoir.
\begin{figure}
\centering
\resizebox{0.40\textwidth}{!}{%
\includegraphics{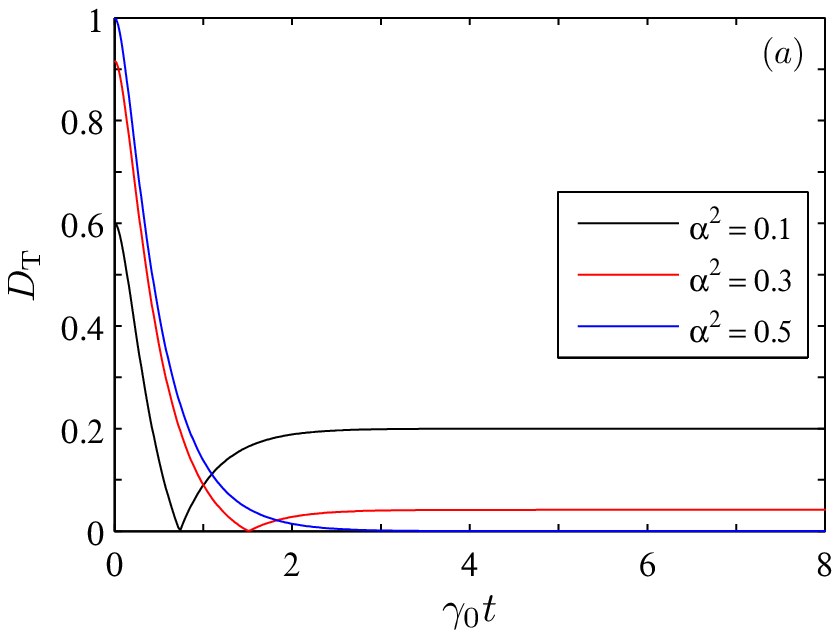}}
\centering
\resizebox{0.40\textwidth}{!}{%
\includegraphics{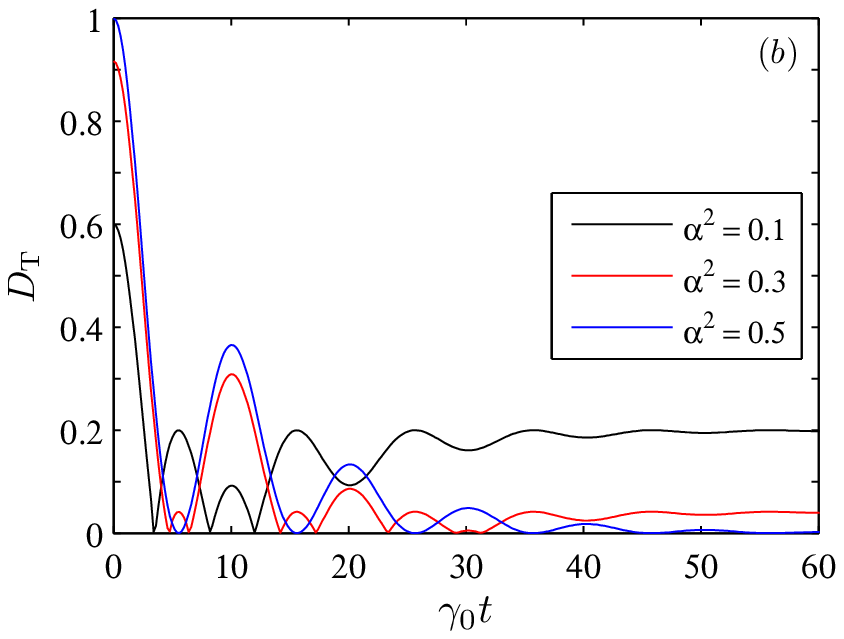}}
\caption{(Color online) Trace distance discord $D_{\rm T}(\rho)$
         versus $\gamma_0 t$ for the initial states $|\Phi\rangle$
         in common reservoir, where the parameter $\lambda$ is
         chosen to be $\lambda=10\gamma_0$ (a) and
         $\lambda=0.1\gamma_0$ (b), respectively.}
         \label{fig:2}
\end{figure}

When the two qubits are subject to the common reservoir, we plotted
in Fig. \ref{fig:2} dynamics of $D_{\rm T}(\rho)$ versus $\gamma_0
t$ for the initial state $|\Phi\rangle$. Different from that of the
independent reservoirs, $D_{\rm T}(\rho)$ here does not behave as
monotonic functions of $|\alpha^2-0.5|$. In the Markovian regime as
shown in Fig. \ref{fig:2}(a), $D_{\rm T}(\rho)$ decays
asymptotically to zero for $\alpha^2=0.5$, and for other $\alpha^2$,
they first decay to the minimum $0$, and then turn out to be
increased to certain steady-state values in the infinite-time limit.
In the non-Markovian regime, as can be seen from Fig.
\ref{fig:2}(b), $D_{\rm T}(\rho)$ oscillates with damped amplitudes
for $\alpha^2=0.5$, and exhibits very complicated behaviors for
other values of $\alpha^2$, which are induced by the combined
effects of the non-Markovianity and the reservoir-mediated
interaction between the two qubits. But in the long-time limit, the
reservoir-mediated interaction between the two qubits dominates, and
$D_{\rm T}(\rho)$ arrives at steady-state values which are
completely the same as those for the Markovian case in Fig.
\ref{fig:2}(a).

\begin{figure}
\centering
\resizebox{0.40\textwidth}{!}{%
\includegraphics{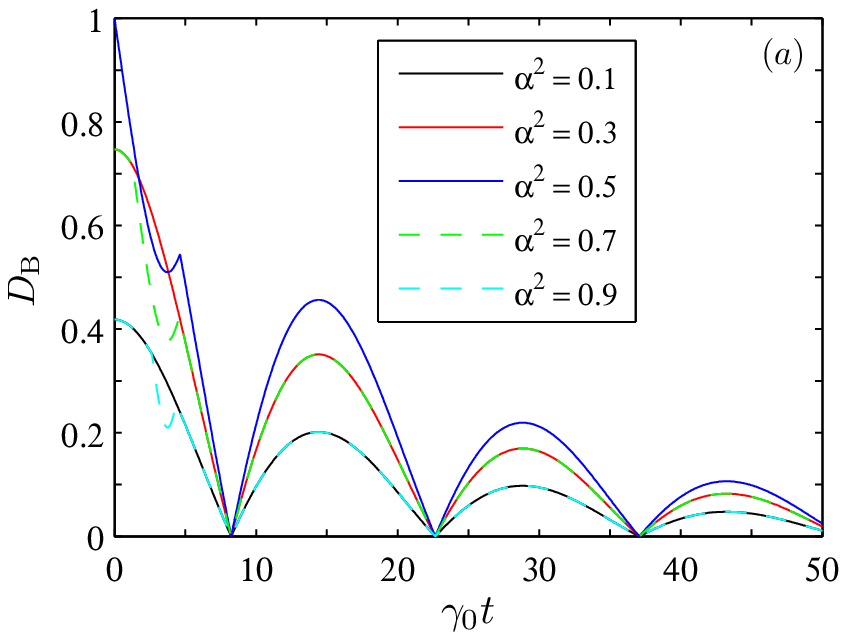}}
\centering
\resizebox{0.40\textwidth}{!}{%
\includegraphics{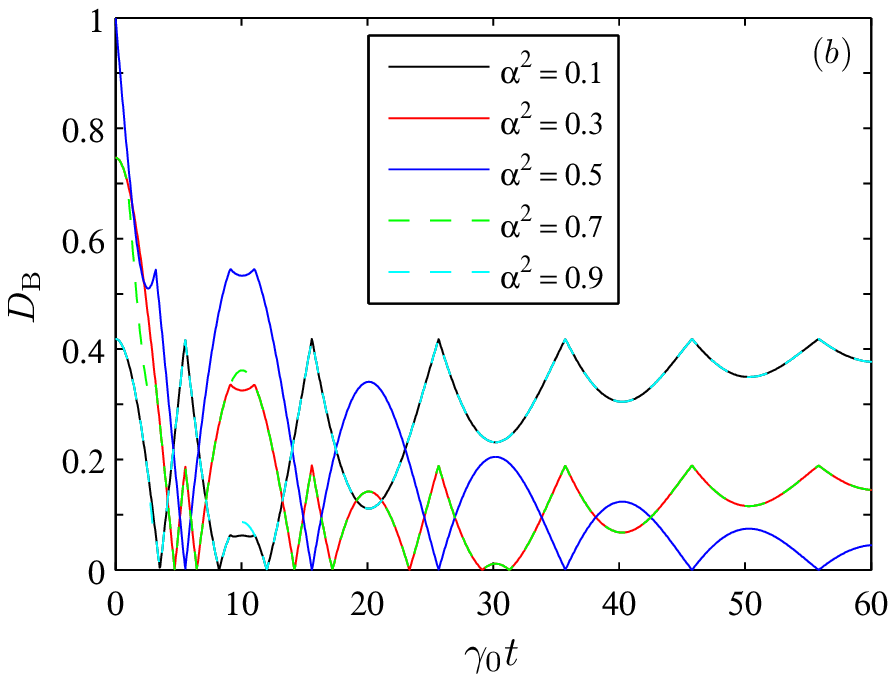}}
\caption{(Color online) Bures distance discord $D_{\rm B}(\rho)$
         versus $\gamma_0 t$ for the initial states $|\Phi\rangle$ in
         independent reservoirs (a) and common reservoir (b),
         both with $\lambda=0.1\gamma_0$.}
         \label{fig:3}
\end{figure}

We now turn to discuss robustness of the Bures distance discord
under the system-environment coupling. Due to the different distance
measures adopted, one may expect that $D_{\rm B}(\rho)$ and $D_{\rm
T}(\rho)$ will exhibit different behaviors. In Fig. \ref{fig:3}, we
showed plots of $D_{\rm B}(\rho)$ versus $\gamma_0 t$ for the
initial state $|\Phi\rangle$ in the non-Markovian regime. First, one
can note that $D_{\rm B}(\rho)$ is no longer a symmetric quantity
with respect to $\alpha^2=0.5$, and this is a difference between
$D_{\rm B}(\rho)$ and $D_{\rm T}(\rho)$. But after a critical point
$\gamma_0 t_c$ which is determined by $\alpha^2$ and the
system-environment coupling parameters, the curves for $D_{\rm
B}(\rho)$ with $\alpha^2$ and $|\alpha^2-1|$ converge. Moreover, one
can see that for the case of independent reservoirs, $D_{\rm
B}(\rho)$ behaves as damped oscillations when $t> t_c$, and
disappears instantaneously at the same critical time $t_n$ as that
for $D_{\rm T}(\rho)$.

For the case of common reservoir, as can be seen from Fig.
\ref{fig:3}(b), the memory effects of the reservoir and the
reservoir-mediated interaction between the considering qubits
together induce more complicated behaviors of $D_{\rm B}(\rho)$ than
that for the independent reservoirs. But beyond the short $\gamma_0
t$ region, $D_{\rm B}(\rho)$ oscillates with fixed periods which are
independent of the values of $\alpha^2$. Particularly, one can note
that apart from the special case of $\alpha^2=0.5$ for which $D_{\rm
B}(\rho)$ behaves as damped oscillations and disappears when
$\gamma_0 t\rightarrow \infty$, the peak values $D_{\rm
B}^{\text{peak}}(\rho)$ for the initial states $|\Phi\rangle$ with
other $\alpha^2$ remain unchanged during their time evolution
process, and these peak values equal to their steady-state values
$D_{\rm B}^{\text{stead}}(\rho)$ in the infinite-time limit, at
which the indirect interaction between the two qubits induced by
their simultaneous interactions with the common reservoir dominates.
\begin{figure}
\centering
\resizebox{0.40\textwidth}{!}{%
\includegraphics{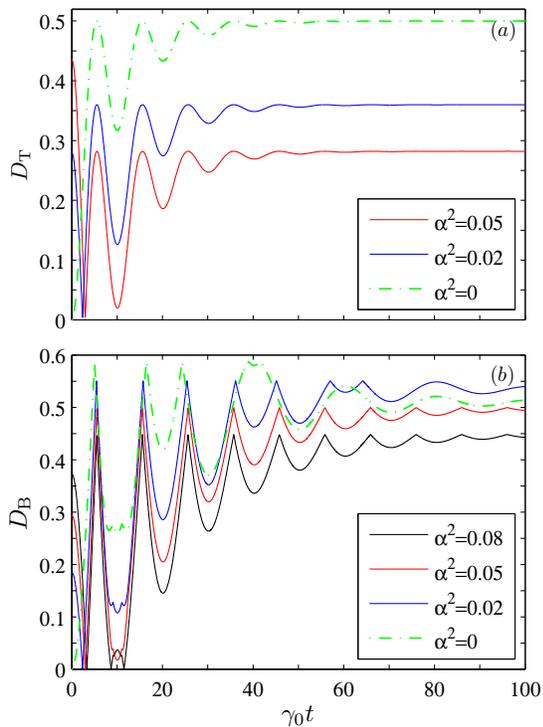}}
\caption{(Color online) Trace distance discord $D_{\rm T}(\rho)$ and
         Bures distance discord $D_{\rm B}(\rho)$ versus $\gamma_0 t$
         for the initial states $|\Phi\rangle$ in common reservoir with
         small values of $\alpha^2$ and $\lambda=0.1\gamma_0$.}
         \label{fig:4}
\end{figure}

Another phenomenon needs to be pay attention to is that for the
cases of $\alpha^2=0.1$ and $0.9$ as represented by the black and
cyan curves in Fig. \ref{fig:3}(b), the steady-state values $D_{\rm
B}^{\text{stead}}(\rho)$ are nearly the same as those for the
initial states $|\Phi\rangle$. Meanwhile, as these steady-state
values are increased with increasing values of $|\alpha^2-0.5|$, one
thus expects naturally that for very small or very large $\alpha^2$,
the Bures distance discord may be enhanced and turns out to be
larger than its initial value. This is indeed the case not only for
the Bures distance discord, but also for the trace distance discord.

As exemplified plots, we illustrated in Fig. \ref{fig:4} the
$\gamma_0 t$ dependence of $D_{\rm T}(\rho)$ and $D_{\rm B}(\rho)$
with various values of $\alpha^2<0.1$. When considering the trace
distance discord, our numerical results show that if
$\alpha^2\lesssim 0.0286$ [e.g., the blue curve for $\alpha^2 =
0.02$ in Fig. \ref{fig:4}(a)], the steady-state value of $D_{\rm
T}(\rho)$ becomes larger than its initial value, and for the special
case of $\alpha^2=0$ which corresponds to the classical state
$|\Phi\rangle=|10\rangle$, the noisy effects of the common reservoir
can even generate trace distance discord, with its maximum be of
about $0.5$. We point out here that the experimental generation of
QD for two ionic qubits via noisy processes has been reported in a
very recent work \cite{generate}, while for generating QD by local
operations, a general approach and powerful result is in Ref.
\cite{luos4}, where it is proved that any separable but quantum
correlated states can be generated from classical states in higher
dimensions via local tracing.

For the Bures distance discord, as can be seen from Fig.
\ref{fig:4}(b), the peak values $D_{\rm B}^{\text{peak}}(\rho)$ for
$\alpha^2\neq 0$ equal to their steady-state values $D_{\rm
B}^{\text{stead}}(\rho)$ and are larger than their initial values.
When $\alpha^2 = 0$, as illustrated by the green dash-dotted curve
in Fig. \ref{fig:4}(b) which corresponds to vanishing $D_{\rm
B}(\rho)$ at the initial time, the noisy process of the common
reservoir can also generate Bures distance discord. For the initial
state $|\Phi\rangle=|10\rangle$, the maximum of $D_{\rm B}(\rho)$ is
of about $0.588$, while its steady-state value is of about $0.495$,
and these are achieved within the scaled time interval $\gamma_0
t\in [0,1000]$. These phenomena show that for certain family of the
initial states, the distance measures of geometric discord can be
preserved well, or even be improved and generated, and it might
indicate a robust pathway to quantum protocols based on them.
\begin{figure}
\centering
\resizebox{0.40\textwidth}{!}{%
\includegraphics{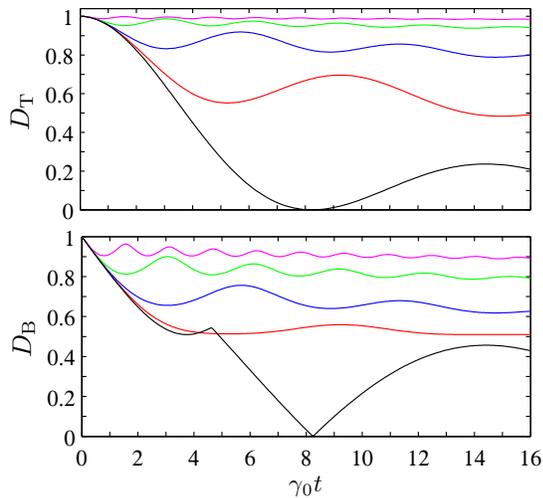}}
\caption{(Color online) Preservation of $D_{\rm T}(\rho)$
         and $D_{\rm B}(\rho)$ in the non-Markovian regime of the
         independent reservoirs. The initial state is $|\Phi\rangle$
         with the parameters $\alpha^2=0.5$, $\lambda=0.1\gamma_0$,
         and the curves from bottom to top correspond to
         $\delta/\gamma_0=0, 0.5, 1, 2$, and $4$.}
         \label{fig:5}
\end{figure}

After making clear their dynamical behaviors, we now try to provide
possible methods for preserving the trace distance and the Bures
distance discord. Of course, for certain family of the initial
states, they can be preserved well or even be enhanced by the common
reservoir, and thus no other specific manipulations are needed for
our purpose. Here, we will further show that for the case of
independent reservoirs, the geometric discord of both $D_{\rm
T}(\rho)$ and $D_{\rm B}(\rho)$ can also be preserved well by the
following strategy.

To be explicitly, we introduce a detuning to the transition
frequency $\omega_0$ by an amount $\delta$, i.e., we replace the
parameter $\omega_0$ in Eq. \eqref{eq10} with the central frequency
$\omega_c=\omega_0-\delta$, and show that it can serve as an
efficient parameter for tuning both $D_{\rm T}(\rho)$ and $D_{\rm
B}(\rho)$. To this end, we displayed in Fig. \ref{fig:5} two
exemplified plots for $D_{\rm T}(\rho)$ and $D_{\rm B}(\rho)$ versus
$\gamma_0 t$, from which one can note that both of them can be
enhanced by introducing detuning, and for the chosen parameters in
Fig. \ref{fig:5}, they begin to oscillate weakly around their
initial values when $\delta=4\gamma_0$. Therefore, it is reasonable
to conjecture that for the case of very large detuning, both $D_{\rm
T}(\rho)$ and $D_{\rm B}(\rho)$ will maintain their initial values
during the time evolution process, and the two kinds of geometric
discord are thus frozen. Such frozen discord provides useful
resource for future operations for quantum protocols relied on them.

\section{Summary}\label{sec:5}
In summary, we have investigated robustness of the trace distance
and the Bures distance geometric discord against decoherence. By
subjecting the considered qubits to structured reservoirs with the
Lorentzian spectral density, we showed that for certain family of
the initial states, the two well-defined geometric discord can be
preserved or improved in the common reservoir. Particularly, the
noisy process induced by the common reservoir can even generate
geometric discord from the classical states. Moreover, we showed
that by introducing detuning to the transition frequency $\omega_0$
of the qubits, an efficient monitoring and long-time preservation of
the geometric discord in non-Markovian independent reservoirs is
also possible, and this inherent robustness might indicate a pathway
to quantum protocols for the open quantum system.

While these two distance measures of geometric discord are well
defined and have important conceptual implications
\cite{trace,bures,Montealegre}, they may exhibit remarkable
features, such as the long-time preservation, improvements, and
generation in the noisy environments as revealed in this paper, and
these make them also important for certain quantum tasks, which
represents a significant challenge and remains as a direction for
future research.

\section*{ACKNOWLEDGMENTS}
This work was supported by NSFC (11205121), NSF of Shaanxi Province
(2010JM1011), and the Scientific Research Program of the Education
Department of Shaanxi Provincial Government (12JK0986).

\newcommand{\PRL}{Phys. Rev. Lett. }
\newcommand{\RMP}{Rev. Mod. Phys. }
\newcommand{\PRA}{Phys. Rev. A }
\newcommand{\PRB}{Phys. Rev. B }
\newcommand{\PRE}{Phys. Rev. E }
\newcommand{\NJP}{New J. Phys. }
\newcommand{\JPA}{J. Phys. A }
\newcommand{\JPB}{J. Phys. B }
\newcommand{\PLA}{Phys. Lett. A }
\newcommand{\NP}{Nat. Phys. }
\newcommand{\NC}{Nat. Commun. }
%

%

\end{document}